\documentclass[
  aps, prl, twocolumn, superscriptaddress, 
  letterpaper, nofootinbib, superscriptcite 
]{revtex4-2}

\usepackage{microtype} 
\usepackage{ragged2e} 
\usepackage{newfloat} 
\usepackage{graphicx, bm, amssymb, xcolor, amsmath, ulem}
\usepackage{caption} 
\usepackage{lineno}

\usepackage[colorlinks=true,
citecolor=blue,    
urlcolor=blue,     
linkcolor=black,
breaklinks=true]{hyperref}

\DeclareCaptionFormat{justified-caption}{%
  \textbf{#1}#2\justifying #3\par%
}
\begin{document}

\title{Topological antilaser}

\author{Rui-Chang Shen}
\affiliation{Department of Physics, The Chinese University of Hong Kong, Shatin, Hong Kong SAR, China}

\author{Chunquan Peng}
\affiliation{National Engineering Research Center of Electromagnetic Radiation Control Materials, University of Electronic Science and
Technology of China, Chengdu 611731, China}
\affiliation{Key Laboratory of Multi-spectral Absorbing Materials and Structures of Ministry of Education, University of Electronic Science and Technology of China, Chengdu 611731, China}
 
 \author{Bingbing Wang}
\affiliation{Department of Physics, The Chinese University of Hong Kong, Shatin, Hong Kong SAR, China}

\author{Wentao Xie}
\affiliation{Department of Physics, The Chinese University of Hong Kong, Shatin, Hong Kong SAR, China}

\author{Siyuan Zhang}
\affiliation{Department of Physics, The Chinese University of Hong Kong, Shatin, Hong Kong SAR, China}

\author{Peiheng Zhou}
\affiliation{National Engineering Research Center of Electromagnetic Radiation Control Materials, University of Electronic Science and
Technology of China, Chengdu 611731, China}
\affiliation{Key Laboratory of Multi-spectral Absorbing Materials and Structures of Ministry of Education, University of Electronic Science and Technology of China, Chengdu 611731, China}

\author{Baile Zhang}
\email{blzhang@ntu.edu.sg}
\affiliation{Division of Physics and Applied Physics, School of Physical and Mathematical Sciences, Nanyang Technological University,
	Singapore 637371, Singapore}
\affiliation{Centre for Disruptive Photonic Technologies, Nanyang Technological University, Singapore 637371, Singapore}

\author{Y. D. Chong}
\email{yidong@ntu.edu.sg}
\affiliation{Division of Physics and Applied Physics, School of Physical and Mathematical Sciences, Nanyang Technological University,
	Singapore 637371, Singapore}
\affiliation{Centre for Disruptive Photonic Technologies, Nanyang Technological University, Singapore 637371, Singapore}

\author{Haoran Xue}
\email{haoranxue@cuhk.edu.hk}
\affiliation{Department of Physics, The Chinese University of Hong Kong, Shatin, Hong Kong SAR, China}
\affiliation{State Key Laboratory of Quantum Information Technologies and Materials, The Chinese University of Hong Kong, Shatin, Hong Kong SAR, China}

\begin{abstract}
Coherent perfect absorption (CPA)—the time-reversed operation of lasing at threshold—relies on finely tuned interference and is intrinsically fragile to disorder and structural imperfections. Whether absorption can be endowed with topological protection, by analogy to topological lasing, has remained an open question. Here, we experimentally demonstrate a topological antilaser: the time-reversed counterpart of a topological laser, in which chiral edge modes of a photonic lattice enable perfect light absorption protected by topology. Using a nonreciprocal microwave network with low intrinsic loss, we show that the topological antilaser preserves near-unity absorption under strong disorder, and—unlike conventional antilasers—remains functional for arbitrary placements of dissipation and input ports, even when the lattice is strongly perturbed. This robustness arises from the disorder-immune propagation and stable spatial profile of the topological edge modes. Our results establish topologically protected absorption as the missing counterpart of topological lasing, opening new directions for studying robust energy dissipation, wave control, and coherent-absorption–based detection technologies.

\end{abstract}

\maketitle

\noindent\textbf{Introduction}\\
Since the discovery of the integer quantum Hall effect~\cite{klitzing1980new}, topological physics has reshaped our understanding of wave phenomena by revealing that global quantized invariants like the Chern number can endow physical observables with extraordinary robustness~\cite{hasan2010colloquium, qi2011topological}. Transplanting these ideas into photonics enabled topological photonic insulators, in which chiral edge modes can propagate unidirectionally along the boundary and remain immune to backscattering from strong disorder~\cite{haldane2008possible, wang2008reflection, wang2009observation,hafezi2011robust, rechtsman2013photonic, khanikaev2013photonic, wu2015scheme, gao2018topologically, lu2014topological, ozawa2019topological}. This robustness has recently been exploited in topological lasers, novel devices in which optical gain applied to a topological mode stabilizes single-mode lasing, enhances slope efficiency, and suppresses sensitivity to fabrication imperfections~\cite{st2017lasing, bahari2017nonreciprocal, harari2018topological, bandres2018topological,zeng2020electrically, kim2020multipolar, dikopoltsev2021topological, yang2022topological}.

In contrast to light generation, light absorption has traditionally been regarded as a local, dissipative process with no meaningful connection to topology. An extreme form of light absorption is coherent perfect absorption (CPA)~\cite{chong2010coherent,baranov2017coherent}---the time-reversed counterpart of lasing at threshold---in which perfect absorption is achieved through the destructive interference of coherent inputs~\cite{longhi2010pt, chong2011pt, wan2011time, zhang2012controlling, wong2016lasing, wei2014symmetrical, song2014acoustic,roger2015coherent,pichler2019random, wang2021coherent, slobodkin2022massively}. The CPA concept reveals a deep connection between light generation and annihilation, and provides a practical way to achieve strong absorption even in low-loss materials. However, CPA is acutely sensitive to structural imperfections, environmental perturbations, and the precise distribution of loss. The fragility of this phenomenon leads us to the question of whether it could benefit from topological protection.

Here, we bridge these previously unconnected domains by introducing topological principles into the absorption process itself. We propose and experimentally demonstrate a topological antilaser, the direct time-reversed analogue of a topological laser, in which a topologically protected chiral edge mode is driven into a CPA condition. Instead of being amplified by gain, the chiral mode is perfectly absorbed within a dissipative region while retaining the hallmark robustness of topological transport.

\begin{figure*}
\includegraphics[width=\textwidth]{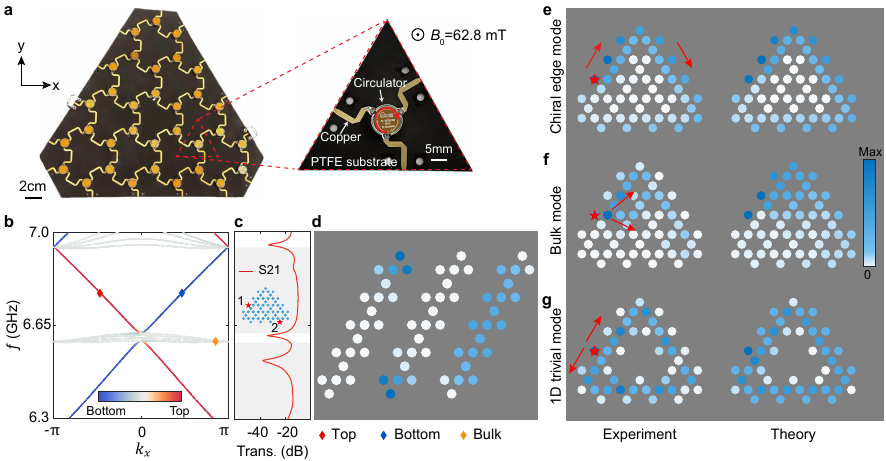}
\caption{\textbf{Scattering network model.} \textbf{a}, Photograph of a fabricated scattering network, with an inset showing the details of a single circulator. \textbf{b}, The dispersion of the scattering network under a slab geometry with periodic boundary conditional along the $x$ direction and five unit cells along the $y$ direction.  \textbf{c}, Experimentally measured edge transmission spectrum, with the shaded regions indicating the bulk bandgaps. The excitation (detection) port position is denoted as port 1(2) in the inset. \textbf{d}, Profiles of the modes highlighted by colored markers in (\textbf{b}). \textbf{e}-\textbf{g}, Measured (left) and calculated (right) field distributions under a single-port excitation for a chiral edge mode at 6.765 GHz (\textbf{e}), a bulk mode at 6.591 GHz (\textbf{f}), and a trivial mode at 6.927 GHz (\textbf{g}).}
\label{fig1}
\end{figure*}

The time-reversed extension of topological lasing leads to two broad consequences, which we investigate experimentally. First, the topological antilaser exhibits enhanced stability against disorder, maintaining near-perfect absorption even when substantial structural perturbations detune conventional antilasers.
Second, the CPA condition can be achieved for arbitrary placements of dissipation and input/output ports, dramatically relaxing the engineering constraints that limit traditional antilasing devices. These capabilities arise from the disorder-immune propagation and spatial profile of topological edge modes, which ensure that the field distribution remains stable under perturbations and therefore preserves the interference condition required for CPA. Our results establish topologically protected absorption as the missing counterpart to topological lasing, extending the unification of topology and non-Hermitian photonics beyond light generation to include robust, topologically stabilized dissipation. This work opens a pathway to disorder-resistant absorption, energy control, and wave-based sensing across photonic and other classical and quantum platforms.

\medskip
\noindent\textbf{Experimental platform}\\
The first challenge in realizing a topological antilaser lies in identifying a suitable topological photonic platform. While antilasers can theoretically be implemented in any lossy wave system, achieving them in complex structures can be challenging due to the increased number of degrees of freedom requiring precise tuning. We employ a recently-developed experimental platform, consisting of a nonreciprocal microwave-frequency network, to implement a topological antilaser~\cite{zhang2021superior,chen2024anomalous}. This system comprises circulators arranged in a honeycomb lattice (Fig.~\ref{fig1}a), with time-reversal symmetry broken by magnetically biased materials in the circulators. The platform features low losses, primarily from the dielectric substrate and circulators, and can be fabricated precisely via printed circuit board manufacturing, making it ideal for implementing a topological antilaser.

\begin{figure*}
\includegraphics[width=\textwidth]{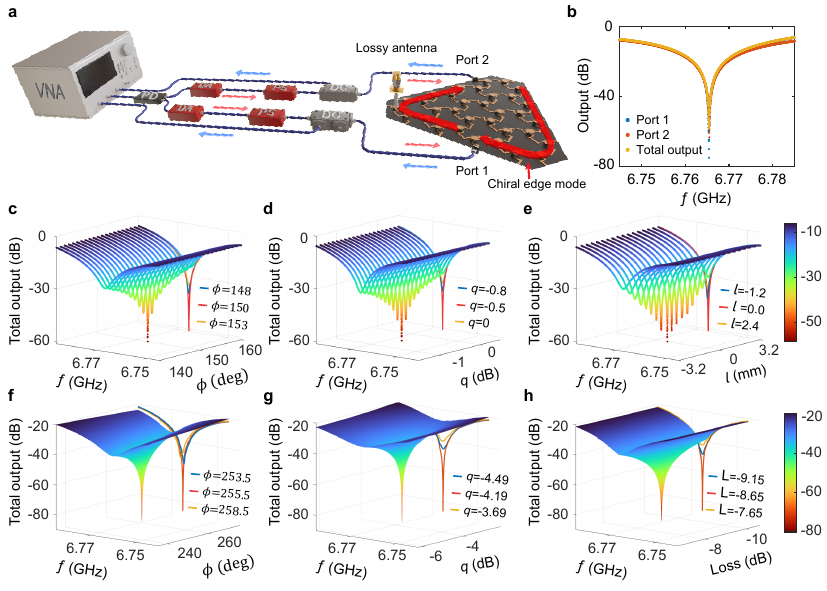}
\caption{\textbf{Antilasing of the chiral edge mode.} \textbf{a}, Schematic of the experimental setup with two input/output ports. \textbf{b}, Measured outputs from port 1 (blue dots) and port 2 (red dots), and the total output (yellow dots). An antilasing is achieved at 6.765 GHz, where the outputs from both ports simultaneously vanish. \textbf{c}-\textbf{e}, Measured total outputs in the $f-\phi$ (\textbf{c}), $f-q$ (\textbf{d}), $f-l$ (\textbf{e}) subspaces, with the antilasing clearly seen in each plot. The line plots on the vertical planes show selected cuts of the corresponding 2D plots. In (\textbf{e}), the position of the lossy antenna corresponding to the antilasing condition is defined as $l=0$ mm. \textbf{f}-\textbf{h}, Calculated total outputs corresponding to (\textbf{c}-\textbf{e}).  In (\textbf{c}-\textbf{h}) , each subplot shows a 2D cross-sections of the 4D parameter space ($f,~\phi,~q,$ and  additional loss), with the two non-scanned parameters fixed at the CPA condition marked by the red lines. }
\label{fig2}
\end{figure*}

The system is depicted in Fig.~\ref{fig1}a, and can be modeled as a scattering network with the circulators acting as the nodes~\cite{zhang2021superior}. Each circulator is connected to three ports and characterized by a 3$\times$3 scattering matrix
\begin{equation}\label{eq1}
\begin{aligned}
S_{0}=\begin{pmatrix}
               R & D & T \\
                T & R & D \\
                D & T & R \\
       \end{pmatrix}
\end{aligned},
\end{equation}
which relates incoming waves to outgoing ones. When translational symmetry is imposed, wave propagation inside this system is described by a Floquet-Bloch equation with the phase of the wave playing the role of quasienergy~\cite{liang2013optical}. The full matrix elements are obtained experimentally by measuring S parameters. Using the measured scattering matrix, we compute the dispersion of this system under a slab geometry (i.e., finite along the $y$ direction and periodic along the $x$ direction)(Supplementary Information section S1). As shown in Fig.~\ref{fig1}b-d, there are multiple topological gaps spanned by chiral edge modes; below, we will use the chiral edge modes around 6.765 GHz (red and blue markers in Fig.~\ref{fig1}b) to construct the topological antilaser. This band structure's topological phase is an anomalous Floquet insulator, which is known to be topologically nontrivial despite the individual bulk bands having zero Chern numbers~\cite{rudner2013anomalous}. However, we emphasize that the phenomenon of topological antilasing is not tied to this specific topological phase; we have verified theoretically that it can also occur in Chern insulators with similarly robust chiral edge modes, like the Haldane and Qi-Wu-Zhang models (Supplementary Information section S5)~\cite{haldane1988model, qi2006topological}.

To confirm the existence of the chiral edge modes, we experimentally measure the field distributions under a single-site excitation at the edge of a triangle-shaped sample. When the driving frequency is inside the topological bandgap (e.g., 6.765 GHz), the measured field is localized around the edges (Fig.~\ref{fig1}e). When the drive is switched to 6.591 GHz, which coincides with a narrow bulk band (yellow marker in Fig.~\ref{fig1}b), the excited field penetrates noticeably into the bulk (Fig.~\ref{fig1}f). As a further comparison,  we fabricate a hollowed structure without magnetic bias, i.e., the time-reversal symmetry is preserved. Here, the hollowed geometry is employed to emulate a 1D trivial edge mode that is strongly confined to the boundary, offering a strict geometric counterpart to the topological edge mode.   Comparing the field distribution in this sample (Fig.~\ref{fig1}g) with the topological one, we observe that the measured amplitudes in the hollowed sample have similar values on either side of the excitation, indicating that wave propagation is bi-directional, as expected.

\begin{figure*}
\includegraphics[width=\textwidth]{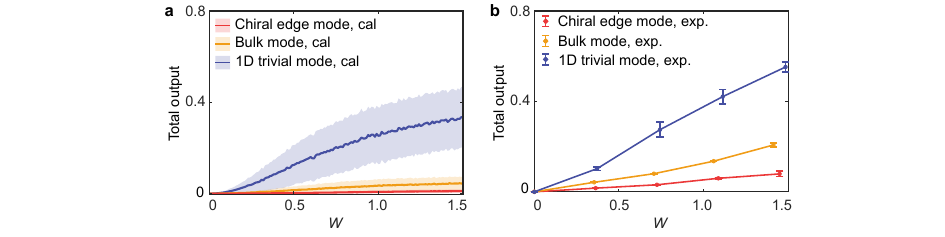}
\caption{\textbf{Unexpected robustness of topological antilaser.} \textbf{a},\textbf{b}, Calculated and measured total outputs versus the disorder strength $W$ defined in the main text. The red, yellow, and blue colors respectively represent antilasers operating via chiral edge modes, bulk modes, and 1D trivial modes.  The error bars denote the standard deviations from five repeated measurements and 1000 disorder realizations for experimental and numerical data, respectively. }
\label{fig3}
\end{figure*}

\medskip
\noindent\textbf{Realizing antilasing via chiral edge mode}\\
We then construct an antilaser by mounting two input/output ports to the edges of the sample, as illustrated in Fig.~\ref{fig2}a. The input signals in the two ports are created from a single port of a vector network analyzer (VNA) by a power divider (PD). A digital attenuator (DA) and a phase shifter (PS) are added to each of the input routes, allowing for the full control of the input signals to reach the antilasing condition. The reflected signal from each port is delivered back to the VNA via a directional coupler (DC), which directs the signals to distinct paths to avoid crosstalk~\cite{zhang2017observation}. Optionally, a microwave antenna with a 50-$\Omega$ resistance can be inserted into a selected site in the lattice to introduce an extra loss~\cite{pichler2019random} (see Methods for details). Under this setup, we can accurately locate antilasing behavior within a four-dimensional parameter space spanned by frequency ($f$), the phase difference between the two input signals ($\phi$), the attenuator difference between the two input signals ($q$), and the additional loss induced by the antenna. The first three parameters can be directly read from the corresponding components, whereas the additional loss is characterized by the inserted length of the antenna inside the sample ($l$).

In our experiment, we scan the four-dimensional parameter space with a dense mesh ($\delta f= 10~{\rm kHz}$, $\delta\phi=1~ {\rm deg}$, $\delta q= 0.1~ {\rm dB}$, and $\delta l= 0.4~ {\rm mm}$).  For now, we fix the lossy antenna to the position indicated in Fig.~\ref{fig2}a, the effects of varying this position will be studied later. An antilaser is identified when the outputs from both ports vanish simultaneously (Fig.~\ref{fig2}b). Figure~\ref{fig2}c-e plot the measured outputs in three different two-dimensional subspaces, clearly revealing antilasing occurring at 6.765 GHz, which as we have noted lies in the topological bandgap. These measured data closely match numerical simulation results (Fig.~\ref{fig2}f-h) (See Supplementary Information section S2 for numerical details). They also exhibit distinctive antilaser features, including the fact that perfect absorption occurs at isolated (``critical'') points in parameter space, as well as the Lorentzian lineshape of the spectrum \cite{chong2010coherent, baranov2017coherent} (See Supplementary Information section S4 for the toy model of the topological antilaser).

\begin{figure*}
\includegraphics[width=0.8\textwidth]{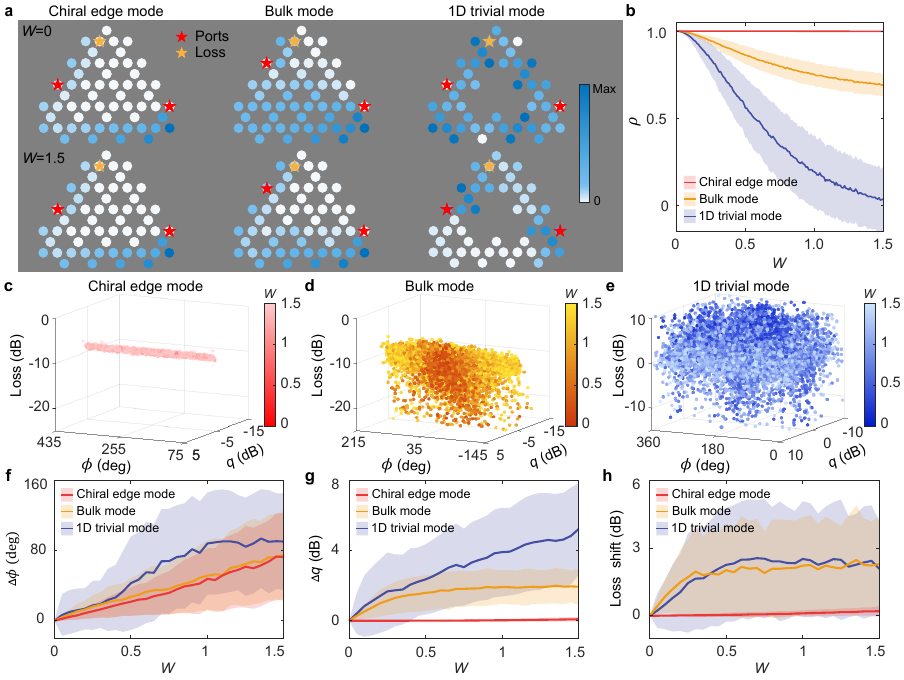}
\caption{\textbf{Analysis of topological antilaser.}  \textbf{a}, Calculated field distributions for the three cases at $W=0$ (upper panels) and at $W=1.5$ (lower panels). The disorder strength $W$ is evaluated at 7 GHz, and the disorder realization is the same as in the experiments. \textbf{b}, The Pearson correlation $\rho$ of the field distribution under CPA conditions as a function of disorder strength.  \textbf{c}-\textbf{e}, Plots of the CPA positions in the parameter space with increasing disorder strength. Each data point represents a parameter set that enables CPA, with color indicating the disorder strength. For each disorder strength, the results are obtained by identifying valid CPA parameter sets across 1000 randomly disordered configurations. For the trivial system, negative loss is required in some cases to realize CPA, implying the necessity of gain support~\cite{zhao2025virtual}. This behavior  can be attributed to the enhanced reflection and suppressed transmission induced by disorder. \textbf{f}-\textbf{h}, Plots of average CPA condition shifts ($\Delta\phi,~\Delta q,$ and loss shift) as functions of disorder strength. }
\label{fig4}
\end{figure*}

\medskip
\noindent\textbf{Unexpected robustness of topological antilaser}\\
Since the antilaser is accessed by fine-tuning, its performance may be seriously affected by random disorder in the lattice. To investigate this, we subject each metallic strip connecting the circulators to an independent random length variation $\Delta L_i$, uniformly distributed between $[-A/2, A/2]$. This induces a random change to the phase of the electromagnetic wave when it passes through the strip. The disorder strength (i.e., the scale of the disorder-induced phase changes) is characterized by $W=2\pi A f \varepsilon^{1/2}_{\rm eff}/c$, where $\varepsilon_{\rm eff}$ is the effective permittivity and $c$ is the speed of light in vacuum.  We fabricate five samples with $A=0, 1.72, 3.45, 5.17, 6.90~{\rm mm}$ , with all the other parameters chosen so that the system would operate at the previously-located antilasing point in the absence of disorder. As shown by the red curves in Fig.~\ref{fig3}a,b, increasing the disorder strength $W$ causes the total output to increase from zero, due to the system becoming detuned from the ideal antilaser condition. However, the total output remains $\lesssim 0.08$ (i.e., absorption remains strong) even after increasing the disorder to a large value ($W=1.45$).

\begin{figure*}
\includegraphics[width=0.8\textwidth]{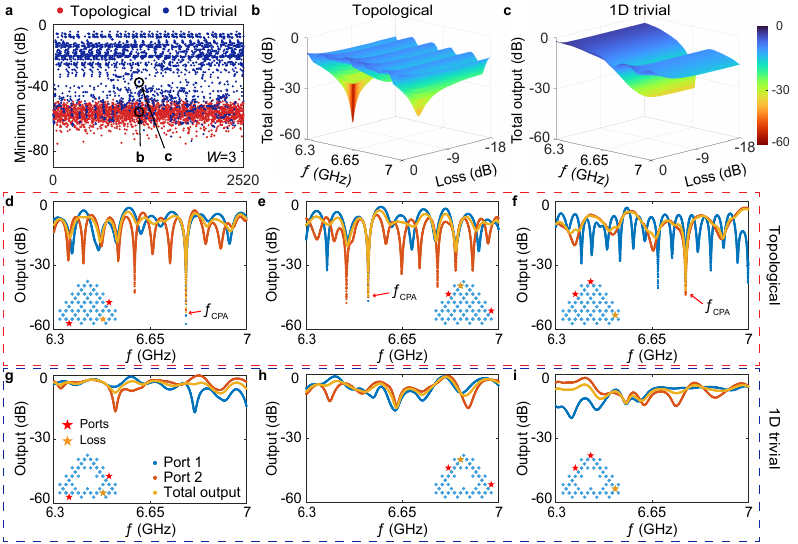}
\caption{\textbf{Topological antilaser in strongly disordered systems.} \textbf{a}, The minimum total output obtained by scanning the four-dimensional parameter space for all possible combinations ($C_{15}^{2}\times 24=2520$) of the port ($C_{15}^{2}$) and lossy site (24) positions in a strongly disordered sample. The red and blue dots represent numerical results for the topological case and the 1D trivial case, respectively. \textbf{b}, An example where antilasing is realized in a disordered topological sample.  \textbf{c}, An example where antilasing is impossible in a disordered trivial sample.  \textbf{d}-\textbf{f}, Experimental realization of antilasing in a strongly disordered topological sample for three distinct port/loss configurations. \textbf{g}-\textbf{i}, Experimentally achieved minimal total outputs in a strongly disordered hollowed sample under the same port/loss configurations. }
\label{fig5}
\end{figure*}

The stability of the topological antilaser is more striking when compared to non-topological ones. We consider two non-topological antilasers based on the configurations previously discussed in Fig.~\ref{fig1}f,g, which respectively utilize (i) modes from the bulk band below the topological bandgap, and (ii) topologically unprotected 1D modes. For the former, we use the same disordered samples as the topological antilaser, with a reduced operating frequency. For each of the non-topological systems, we are able to locate an antilasing condition in the disorder-free regime through parameter tuning. However, we discover that the output grows more rapidly with disorder strength, compared to the topological case, as shown by the blue and yellow curves in Fig.~\ref{fig3}a,b.

\medskip
\noindent\textbf{Analysis of topological antilaser}\\
The robustness of the topological antilaser can be attributed to a distinctive feature of the topologically protected edge modes. In the presence of disorder, the chiral edge modes continue to propagate unidirectionally along the boundary, not subject to Anderson localization~\cite{anderson1958absence} ( Supplementary Information section S3). As a consequence, for a given set of input amplitudes, the field distribution in the lattice does not vary much with disorder, as can be seen from exemplary calculated field distributions for disorder-free and strongly-disordered samples (Fig.~\ref{fig4}a, left panel). When utilizing non-topological modes, however, the field distributions are significantly altered by disorder (Fig.~\ref{fig4}a, middle and right panels). To quantify this effect, we compute the Pearson correlation coefficient of the field distribution,
\begin{equation}
\rho(W) = \frac{\text{cov}(\psi_{\rm W=0}, \psi_{\rm W})}{\sigma_{\psi_{\rm W=0}} \sigma_{\psi_{\rm W}}},    
\end{equation}
where $\psi_{\rm W}$ is the field amplitude under disorder strength $W$, $\text{cov}(X, Y)$ is the covariance between distributions $X$ and $Y$, and $\sigma_{X}$ is the population standard deviation of $X$. The Pearson correlation coefficient $\rho$ approaches 1 when two distributions are linearly correlated. As shown in Fig.~\ref{fig4}b, $\rho$ remains close to 1 with increasing $W$ for the topological case, but experiences sharp drops for the bulk and 1D trivial modes.

A strong correlation indicates that the coherence between the two incident fields is not significantly degraded, implying that parameter combinations realizing CPA remain more stable under weak disorder. Figure~\ref{fig4}c–e show the CPA positions for the three configurations, with each point representing a set that yields antilasing. The results show that disorder-induced shifts of the CPA condition are markedly smaller for the topological edge mode than for the trivial modes.

To further quantify this behavior, we plot the variations of the CPA parameters as a function of disorder strength in Fig.~\ref{fig4}f–h. In our system, disorder is introduced by randomly varying the lengths of metallic strips, thereby altering the propagation phase. The shift in phase difference $\Delta \phi$ shown in Fig.~\ref{fig4}f thus reflects the parameter adjustment required to compensate for disorder-induced phase distortions. Furthermore, Figure~\ref{fig4}g,h demonstrate that the amplitude ratio and additional loss required for CPA in the topological antilaser exhibit substantially enhanced stability against disorder compared with the trivial ones. This provides the direct physical origin for the near-unity absorption sustained by the topological system, as seen in Fig.~\ref{fig3}a,b. Notably, the CPA in our system relies on four tunable parameters ($f,~\phi,~q,$ and additional loss). Since topological edge and bulk modes are confined to distinct frequency ranges, we therefore only study the shifts of the remaining three parameters.

\medskip
\noindent\textbf{Prevalence of topological antilaser under strong disorder}\\
Beyond disorder resistance, topological antilasers dramatically relax the CPA implementation constraints in disordered systems. To see this, we consider a topological sample and a hollowed sample, respectively, both with a disorder strength $W=3$. We examine whether antilasing can be achieved in these two systems. For this purpose, we randomly choose the positions of the two ports and the lossy site on the edges, and scan the whole parameter space to obtain the minimum output that can be reached under each configuration. As shown in Fig.~\ref{fig5}a, the minimum outputs are generally much lower in the topological case. Here, the specific values of the minimum outputs are sensitive to the parameter sweeping steps. Under our numerical setting, we find that -36 dB is a good cutoff line, below which an antilaser is realized for most cases. Based on this criterion, one immediately finds that antilasing is super-universal in the topological case, regardless of the positions of the ports and lossy sites, and even the specific disorder realizations (see Fig.~\ref{fig5}b for an example). In the non-topological case, however, there is a large number of cases (around 73.2\%) where antilasing is impossible (see Fig.~\ref{fig5}c for an example). This property is again a consequence of the fact that the chiral edge mode remains delocalized under disorder and the corresponding mode profile is much more stable than a conventional mode.

Experimentally, we demonstrate this flexibility using three configurations on a $W=3$ disordered sample, as illustrated in Fig.~\ref{fig5}d-f. In all three cases, antilasing can be achieved with properly tuned system parameters in the topological case (Fig.~\ref{fig5}d-f). The situation changes dramatically for a trivial sample with the same positions for the input/output and lossy sites. Figure~\ref{fig5}g-i show the measured output curves that correspond to the minimum total outputs that can be achieved in experiments, which demonstrate the absence of antilasing in the three cases in the trivial sample.

\medskip
\noindent \textbf{Conclusion }\\
In conclusion, we have proposed and experimentally realized a topological antilaser that exploits chiral edge modes to achieve CPA with enhanced robustness.  Due to its topologically protected field stability and homogeneity, the topological antilaser overcomes a key downside of conventional antilasers, i.e., fragility under perturbations. Our experiments have established two kinds of advantage enjoyed by these systems: (1) greater stability against disorder, maintaining low output under perturbations, and (2) relaxed implementation requirements, enabling antilasing in strongly disordered systems without special engineering. These results open new frontiers for both topological photonics and wave absorption. In future, it will be interesting to explore using more types of topological modes, which may bring other forms of robustness to antilasers, involving other features such as lattice symmetries or pseudospin degrees of freedom. Topological antilaser can also be realized in other platforms with controlled system parameters, such as optical topological silicon chips and phononic topological crystals. Apart from Hermitian topological modes, it will also be interesting to study the physics of intrinsic non-Hermitian topological modes (e.g., non-Hermitian skin modes~\cite{yao2018edge}) in antilasers, which may open new possibilities.

\bigskip
\noindent \textbf{\large Methods }\\
\noindent\textbf{Sample details}\\
The nonreciprocal network sample consists of three-port circulators interconnected with waveguides. The circulator (UIYSC9B55T6, UIY Co.) incorporates a samarium-cobalt (SmCo) permanent magnet that generates a bias magnetic field of approximately 62.8 mT to break the time-reversal symmetry. The waveguide is fabricated using Rogers RT/duroid 5880 laminate, which features a three-layer structure comprising $35\,\rm{\mu} \mathrm{m}$  thick copper cladding on both sides with a polytetrafluoroethylene (PTFE) dielectric core. The microstrip lines connecting the circulators were designed with a length ($L$) of 29.8 mm (in the absence of disorder) and a width of 1.65 mm, creating a frequency-dependent phase delay characterized by $\phi=2\pi L f \varepsilon^{1/2}_{\rm eff}/c$, where $\varepsilon_{\rm eff}$ is the effective permittivity of the  microstrip line and $c$ is the speed of light in vacuum. Precision fabrication was achieved by etching with $\pm 20~\rm{\mu} \mathrm{m}$ dimensional accuracy. Circulator components were mounted using surface mount technology (SMT) with reflow soldering. For the trivial samples, a circular waveguide was employed to replace the circulator. The outer diameter of the circular waveguide is $R_1 = 5.4$ mm, while the inner diameter is given by $R_2 = 3.75$ mm.

\bigskip
\noindent\textbf{Experimental measurements}\\
As illustrated in Fig.~\ref{fig2}a, the two input waves are generated from a single port of a four-port vector network analyzer (VNA, ZNB26, R$\&$S) and asubsequently divided using a power splitter to ensure strict frequency synchronization. Before being injected into the sample, each wave passes through a separate phase shifter (LPS-802, Vaunix) and a digital attenuator (LDA-908V, Vaunix), enabling independent control of both phase and amplitude of the two input waves. The phase shifters have a resolution of $1^{\circ}$ while the digital attenuators offer 0.1 dB step precision. Wave injection is achieved through directional couplers, with the coupled port of the directional couplers routing both output signals back to two distinct VNA ports for independent measurements. This configuration allows for simultaneous acquisition of data from both output channels. The lossy antenna is terminated with a 50 ohm impedance and its position is controlled by a three-dimensional positioning stage. The process of loss decrease involves moving the antenna away from the position that satisfies the CPA condition (the zero point). The process of loss increase is achieved by moving the other antenna closer to the waveguide from a starting point 3.2 mm above the zero point.

\bigskip
\noindent\textbf{Data availability} The data presented in this work are available at XXX (URL to be inserted upon publication). 

\bigskip
\noindent\textbf{Code availability} The code used to analyse the data and produce the figures in this article is available at XXX (URL to be inserted upon publication). 

\bigskip
\noindent\textbf{Acknowledgements}
This work was supported by the National Key Research and Development Program of China under Grant No. 2025YFA1412300 (H.X.), the National Natural Science Foundation of China under Grant Nos. 62401491 (H.X.),  52425205 and 52021001 (P.Z.), the Research Grants Council of the Hong Kong Special Administrative Region, China under Grant No. 24304825 (H.X.), the Singapore National Research Foundation (NRF) under competitive Research Program (CRP) Nos. NRF-CRP23-2019-0005, NRF-CRP23-2019-0007, and NRF-CRP29-2022-0003 (Y.C. and B.Z.), the NRF investigatorship No. NRF-NRFI08-2022-0001 (Y.C.),  the Singapore Ministry of Education Academic Research Fund Tier 2 No. MOE-T2EP50123-0007 and Tier 1 Nos. RG139/22 and RG81/23 (B.Z.), and the Guangdong Provincial Quantum Science Strategic Initiative under Grant No.~GDZX2501012 (H.X.) 

\bigskip
\noindent\textbf{Author contributions} B.Z., Y.D.C. and H.X. conceived the idea and supervised the project. R.S. performed the calculations. R.S., P.Z. and H.X. designed the experiment. R.S., C.P., B.W., W.X and S.Z. conducted the experiment. R.S., P.Z., B.Z., Y.D.C. and H.X. analyzed the data. R.S., B.Z., Y.D.C. and H.X. wrote the manuscript. 

\bigskip
\noindent\textbf{Competing interests} The authors declare no competing interests.

\end{document}